\documentclass[prd,nofootinbib,amsmath,superscriptaddress,preprint]{revtex4-1}

\usepackage{color}
\usepackage{xcolor}
\usepackage{graphicx}

\begin{document}

\title{Nonlinear vacuum electrodynamics and spontaneous breaking of Lorentz symmetry}

\author{C. A. Escobar}
\email{carlos\_escobar@fisica.unam.mx}
\affiliation{Instituto de F\'\i sica, Universidad Nacional Aut\'onoma de M\'exico, Apdo.\ Postal 20-364, Ciudad de M\'exico 01000, M\'exico}
\author{R. Potting}
\email{rpotting@ualg.pt}
\affiliation{Departamento de F\'isica, Universidade do Algarve, FCT, 8005-139 Faro, Portugal}
\affiliation{CENTRA, Instituto Superior T\'ecnico, Universidade de Lisboa,
Avenida Rovisco Pais, Lisboa, Portugal}

\begin{abstract}
We study nonlinear vacuum electrodynamics in a first-order formulation
proposed by Pleba\'nski.
By applying a Dirac constraint analysis,
we derive an effective Hamiltonian.
We show that there exists a large class of potentials
for which the effective Hamiltonian is bounded from below,
while at the same time possessing stationary points in which
the field strength acquires a nonzero vacuum expectation value.
The associated spontaneous breaking of Lorentz symmetry
might in principle be detected by coupling the model
to a suitable external current, or to gravity.
We show that the possible vacua can be classified in four classes.
We study some of their properties,
using explicit examples for illustration.
\end{abstract}

\maketitle

\section{Introduction}
Nonlinear electrodynamics ``in vacuum'' is a topic that has received attention
ever since the seminal paper by Born and Infeld \cite{BornInfeld} who were
interested in finding a consistent modification of electrodynamics in
which the energy of the electrostatic potential of a point charge is finite.
In Born-Infeld electrodynamics,
the nonlinear form of the action has the effect of turning the
vacuum into a nontrivial medium,
allowing us to write the modified Maxwell
equations in terms of the displacement and magnetic fields
which can be expressed in terms of the electric field and magnetic induction
through nontrivial Lorentz-covariant field-dependent expressions.
In this paper, our focus of attention is a subclass of models of nonlinear
electrodynamics in which there are nontrivial stationary points that can
serve as vacua of the theory.
In particular, if the field strength in such a stationary point is nonzero,
this can entail the spontaneous breaking of Lorentz symmetry.
Spontaneous symmetry breaking (SSB) of Lorentz invariance is generally
assumed to be at origin of the Standard-Model Extension (SME)
\cite{SME,SMEgravity},
where Lorentz-violating tensor coefficients are assumed to arise from vacuum
expectation values of some basic fields belonging to a more fundamental
underlying model, like string field theory or other models of quantum gravity.
In a class of models in the context of quantum field theory,
which received a lot of attention in the literature,
such a vacuum expectation value is acquired by a vector field $B_\mu$.
The photon then arises as the corresponding Goldstone boson
of the global spontaneous Lorentz symmetry breaking.
The original idea for this goes back to works of
Dirac \cite{Dirac-aether}, Bjorken \cite{Bjorken} and Nambu \cite{Nambu},
who considered a quadratic constraint forcing the vector potential
in electrodynamics to fluctuate around a nonzero vacuum value.
While the latter breaks Lorentz invariance,
no physical Lorentz-violating effects exist,
as the constraint essentially serves as a gauge condition for electrodynamics. 
In the so-called bumblebee models
proposed by Kostelecky et.\ al.\ \cite{KosteleckySamuel,BluhmKostelecky},
the vacuum expectation value for the
vector field is generated by adding an explicit nonderivative potential
designed to break Lorentz symmetry via a nonzero vacuum expectation value
$\langle B_\mu\rangle$.
Note that such a potential breaks also gauge invariance.
The subsequent symmetry breaking splits the original four
degrees of freedom into three vectorial Nambu-Goldstone bosons
satisfying the constraint $B_\mu B^\mu = \pm b^2$,
to be identified with the photon, plus a massive scalar field.
It has been shown, at least at tree level \cite{Nambu} and at one-loop order
\cite{AzatovChkareuli}, that any Lorentz-violating effects in
scattering amplitudes are physically unobservable in the high-mass
limit in which the excitations of the scalar field can be ignored.
Nevertheless, the appearance of the extra degrees of freedom of the vector
field does give rise to certain issues. 
For instance, an extra ``fossile'' (or vacuum) electric current can arise
\cite{ChkareuliFroggattNielsen},
possibly compromising the conservation of the usual current in QED.
The absence of gauge invariance does not protect the form of
the kinetic term anymore from the emergence of non-gauge-invariant
contributions through quantum effects \cite{KrausTomboulis}.
More seriously, in order to assure stability of the model,
it is typically necessary to restrict the phase space to a suitable subspace
\footnote{For vector bumblebee models with a smooth potential, or with a linear Lagrange-multiplier potential,
it has been shown \cite{BluhmGagnePottingVrublevskis}
that the Hamiltonian is bounded from below only for a restricted phase
space, essentially reducing the theory
to Einstein-Maxwell electrodynamics up to possible
SME matter-sector couplings.
See also \cite{BluhmFungKostelecky} for an analysis including gravity.}.
Other models generalizing the idea to models involving
an antisymmetric tensor field \cite{AltschulBaileyKostelecky}
or gravity \cite{KosteleckyPotting} have been proposed as well.

The most important advantage approach of nonlinear electrodynamics
we are exploring in the current work is that it is not the vector potential
but, rather, the field strength that acquires a vacuum expectation value.
This way, gauge invariance is maintained from the beginning,
avoiding the associated problems with its breaking.
In particular, as we will show in this work,
there exists a large class of Lagrangians in which the
effective Hamiltonian is strictly bounded from below,
assuring stability.
Moreover, in many cases the effective
Hamiltonian has nontrivial minima which can serve as
(alternative) vacua for the theory.
In such vacua, the presence of nonzero field strength can give rise to
observable Lorentz-violating effects through the coupling to other fields.
As we will see, the dynamics of the fluctuations around these vacua
is unlike the one of the Maxwell field described by the
usual Lagrangian $-\frac{1}{4}F_{\mu\nu}F^{\mu\nu}$.
Consequently, in the scenarios described in this work
we envision the vector field not to
correspond to the usual Maxwell electrodynamics,
but to some other (so far unobserved) $U(1)$ gauge field.
Rather than following the original approach of Born and Infeld,
who used the field strength and the metric as fundamental variables,
we will use the simpler,
first-order approach pioneered by Pleba\'nski \cite{Plebanski}.
Here the vector potential is added as an independent degree of freedom,
allowing us to adopt a fixed Minkowski metric,
which is sufficient for the purposes of this paper.

A similar approach to nonlinear electrodynamics as a gauge-invariant
way to generate Lorentz-violating effects as the one followed in this paper
was developed by Alfaro and Urrutia \cite{AlfaroUrrutia}.
As a motivation for considering nonlinear electrodynamics they
analyzed in some detail the way effective photon interactions arise
in QED if one integrates out massive gauge bosons and fermions.
We will not delve into this issue in this work,
and refer to \cite{AlfaroUrrutia} for more details.
The main focus of attention in \cite{AlfaroUrrutia} were effective potentials
with local minima.
However, the field configurations around which the expansion are performed
in that work are not local minima of the effective Hamiltonian
we will derive in this work,
and thus it is difficult to envision how they can serve as stable vacua,
in particular when coupled to other degrees of freedom.
Nevertheless, the approach pioneered in \cite{AlfaroUrrutia} certainly
served as an inspiration for the current work.

This paper is organized as follows.
In section \ref{sec:first-order} we review the first-order formulation
of nonlinear electrodynamics used in this work,
presenting the Born-Infeld Lagrangian as an example.
In section \ref{sec:Hamiltonian_analysis} we present a Hamiltonian
analysis of the phase space,
using a Dirac-type analysis to identify first- and second-class constraints
and derive an effective Hamiltonian for the model.
We then analyze various types of stationary points
in section \ref{sec:stability} by using appropriate specific examples.
Finally, we present our conclusions, as well as an outlook,
in section \ref{sec:conclusion}.

\section{First-order formulation of nonlinear electrodynamics}
\label{sec:first-order}
In this section we will review the properties of the Pleba\'nski class of
nonlinear electrodynamics models that we will use in this work.
We start with the first-order action
\begin{equation}
S = \int d^4x\,\mathcal{L}
\label{action}
\end{equation}
in Minkowski space, where the Lagrangian density 
\begin{equation}
\mathcal{L}= - P^{\mu\nu} \partial_\mu A_\nu - V(P,Q) - A_\mu J^\mu
\label{Lagrangian-density}
\end{equation}
depends on the vector potential $A_\mu$ and on the antisymmetric tensor
$P^{\mu\nu}$,
which we treat as independent fields.
The potential $V$ is a scalar function of $P^{\mu\nu}$.
It can be shown \cite{invariants} that it must be a function
of the quadratic invariants
\begin{equation}
P = \frac{1}{4}P_{\mu\nu} P^{\mu\nu}\>
\qquad\mbox{and}\qquad
Q = \frac{1}{4}P_{\mu\nu} \tilde P^{\mu\nu}\>
\label{P-Q}
\end{equation}
where
\begin{equation}
\tilde P^{\mu\nu} = \frac{1}{2} \epsilon^{\mu\nu\rho\sigma} P_{\rho\sigma}\>.
\label{P-tilde}
\end{equation}
Here we define $\epsilon^{0123} = -\epsilon_{0123} = 1$.
In this work we choose the metric convention $(+,-,-,-)$ and use
natural, Heaviside-Lorentz units (with $c = \hbar = 1$).
%
%
The last term in the Lagrangian (\ref{Lagrangian-density})
is a coupling to the external current density $J^\mu$,
which is taken to be conserved:
\begin{equation}
\partial_\mu J^\mu = 0\>.
\label{current-conservation}
\end{equation}
The action (\ref{action}) is then invariant under the gauge transformation
\begin{equation}
A_\mu \to A_\mu + \partial_ \mu \Lambda\>, \quad\quad P^{\mu\nu} \to P^{\mu\nu}
\label{gauge-transformation}
\end{equation}
for arbitrary local gauge parameter $\Lambda$.
The equations of motion of (\ref{action}) read
\begin{eqnarray}
\label{inhomogeneous-Maxwell}
\frac{\delta S}{\delta A_\mu} &=& -\partial_\nu P^{\mu\nu} - J^\mu = 0 \>,\\
\frac{\delta S}{\delta P^{\mu\nu}} &=&
- \frac12(\partial_\mu A_\nu -\partial_\nu A_\mu)
- \frac{1}{2}\left(V_P P_{\mu\nu} + V_Q \tilde P_{\mu\nu}\right) = 0 \>,
\label{Amu-equation}
\end{eqnarray}
where the lower indices on $V$ indicate the partial derivatives
\begin{equation}
V_P = \frac{\partial V}{\partial P}\>,
\qquad V_Q = \frac{\partial V}{\partial Q}\>.
\end{equation}
Introducing the antisymmetric tensor (field strength)
\begin{equation}
F_{\mu\nu} \equiv \partial_\mu A_\nu -\partial_\nu A_\mu\>,
\label{F-A}
\end{equation}
Eq.\ (\ref{Amu-equation}) yields the constitutive relation
\begin{equation}
F_{\mu\nu} = -2\frac{\partial V}{\partial P^{\mu\nu}}
= - V_P P_{\mu\nu} - V_Q \tilde P_{\mu\nu}\>.
\label{constitutive-relation1}
\end{equation}
From definition (\ref{F-A}) we obtain the Bianchi identities
\begin{equation}
\partial_\mu \tilde{F}^{\mu\nu} = \frac{1}{2}\partial_\mu\epsilon^{\mu\nu\rho\sigma} F_{\rho\sigma} = 0\>.
\label{homogenous-Maxwell}
\end{equation}
The constitutive relation (\ref{constitutive-relation1}) can be inverted
by considering $\mathcal{L}$ to be a function of $F^{\mu\nu}$ (as well as
$A_\mu$ and $J^\mu$). 
By Lorentz invariance, $\mathcal{L}$ should then be a function of the
invariants \cite{invariants}
\begin{equation}
F = \frac{1}{4}F_{\mu\nu} F^{\mu\nu}\>
\qquad\mbox{and}\qquad
G = \frac{1}{4}F_{\mu\nu} \tilde F^{\mu\nu}\>.
\label{F-G}
\end{equation}
For arbitrary variation of the fields it follows
\begin{eqnarray}
\delta\mathcal{L} &=& -\frac{1}{2}\delta P^{\mu\nu}F_{\mu\nu}-\frac{1}{2}P^{\mu\nu}\delta F_{\mu\nu} - \frac{\partial V}{\partial P^{\mu\nu}}\delta P^{\mu\nu}
-\delta A_\mu J^\mu \nonumber\\
&=& -\frac{1}{2}\left(F_{\mu\nu} + 2\frac{\partial V}{\partial P^{\mu\nu}} \right)
\delta P^{\mu\nu} - \frac{1}{2}P^{\mu\nu} \delta F_{\mu\nu} - \delta A_\mu J^\mu \nonumber\\
&=& -\frac{1}{2} P^{\mu\nu} \delta F_{\mu\nu} - \delta A_\mu J^\mu\>,
\label{variation-L}
\end{eqnarray}
where the last equality follows from Eq.\ (\ref{constitutive-relation1}).
On the other hand, we have
\begin{equation}
\delta\mathcal{L} = \frac{1}{2}\left(\mathcal{L}_F F^{\mu\nu} + \mathcal{L}_G \tilde F^{\mu\nu}\right) \delta F_{\mu\nu} - \delta A_\mu J^\mu
\end{equation}
so that 
\begin{equation}
P^{\mu\nu} = -\mathcal{L}_F F^{\mu\nu} - \mathcal{L}_G \tilde F^{\mu\nu}
\label{constitutive-relation2}
\end{equation}
which is the inverse of the constitutive relation (\ref{constitutive-relation1}).

The above relations can be expressed in terms
of the vector fields $\vec D$, $\vec E$, $\vec H$ and $\vec B$ by
defining
\begin{equation}
P^{\mu\nu} =\left(
\begin{array}{c c c c}
0   & -D_x & -D_y & -D_z\\
D_x & 0    & -H_z & H_y\\
D_y & H_z  & 0    & -H_x\\
D_z & -H_y & H_x  & 0 
\end{array}\right)
\end{equation}
and
\begin{equation}
F^{\mu\nu} =\left(
\begin{array}{c c c c}
0    &  -E_x &  -E_y &  -E_z\\
E_x & 0    & -B_z & B_y\\
E_y & B_z  & 0    & -B_x\\
E_z & -B_y & B_x  & 0 
\end{array}\right),
\end{equation}
or, equivalently,
\begin{eqnarray}
D_i &&\equiv P_{0i} = (D_x, D_y, D_z)\>, \\
E_i &&\equiv F_{0i} = (E_x, E_y, E_z)\>, \\
H_i &&\equiv \tilde P_{0i} = (H_x, H_y, H_z)\>, \\
B_i &&\equiv \tilde F_{0i} = (B_x, B_y, B_z)\>.
\label{field-defs}
\end{eqnarray}
The invariants $P$, $Q$, $F$ and $G$ then become
\begin{eqnarray}
P &&= \frac{1}{2}(\vec{H}^2 - \vec{D}^2)
\qquad \mbox{and} \qquad
Q = -\vec H \cdot \vec D\>,
\label{PQ-HD} \\
F &&= \frac{1}{2}(\vec B^2 - \vec E^2)
\qquad \mbox{and} \qquad
G = -\vec B \cdot \vec E
\label{FG-BE}
\end{eqnarray}
while the constitutive relations (\ref{constitutive-relation1}) and
(\ref{constitutive-relation2}) can be expressed as:
\begin{equation}
\left(\begin{array}{c}\vec{E} \\ \vec{B} \end{array}\right) =\left(\begin{array}{c c}-V_P & -V_Q \\ V_Q & -V_P \end{array}\right)\left(\begin{array}{c}\vec D \\ \vec H \end{array}\right)
\label{constitutive-relations3a}
\end{equation}
and
\begin{equation}
\left(\begin{array}{c}\vec D \\ \vec H \end{array}\right) =
\left(\begin{array}{c c}-\mathcal{L}_F & -\mathcal{L}_G \\ \mathcal{L}_G & -\mathcal{L}_F \end{array}\right)
\left(\begin{array}{c}\vec E \\ \vec B \end{array}\right)\>.
\label{constitutive-relations3b}
\end{equation}
These generalize the usual relations
$\vec D = \epsilon \vec E$ and $\vec B = \mu \vec H$.
Equations (\ref{inhomogeneous-Maxwell}) and (\ref{homogenous-Maxwell})
then take the familiar form 
\begin{eqnarray}
\vec\nabla \cdot \vec D &=& J^0 \>,\label{Gauss-D}\\
\vec\nabla \times \vec H -\frac{\partial\vec D}{\partial t} &=& \vec J\>, \label{Ampere}\\
\vec\nabla \cdot \vec B &=& 0\>, \label{Gauss-B}\\
\vec\nabla \times \vec E + \frac{\partial\vec B}{\partial t} &=& 0 \label{Faraday}
\end{eqnarray}
of the Maxwell equations in a material medium.
For a detailed treatment of the Pleba\'nski formalism, see \cite{Plebanski}.

\section{Hamiltonian analysis}
\label{sec:Hamiltonian_analysis}

Having presented in the previous section
the Pleba\'nski approach to nonlinear electrodynamics (PNLED)
in a Lagrangian framework, we will now subject it to a Hamiltonian analysis.
This will allow us to address important issues such as stability and the
possible existence of nontrivial local minima.
Naively, one might expect that these questions can be addressed by analyzing
the potential $V(P,Q)$ introduced in the previous section.
However, we will see in the following that this is not the case for the
action (\ref{action}),
and that the relevant functional is in fact the Hamiltonian density itself.

We start by writing Lagrange density (\ref{Lagrangian-density}) as 
\begin{equation}
\mathcal{L}=  A_i\partial_0 P^{0i} - (\partial_i A_\nu) P^{i\nu}- V(P,Q) - A_\mu J^{\mu}
\label{Lagrangian-density2}
\end{equation}
where we performed a partial integration.
Defining the canonical momenta
\begin{equation}
\Pi_0^A = \frac{\partial \mathcal{L}}{\partial \dot{A_0}}\>,\quad
\Pi_i^A = \frac{\partial \mathcal{L}}{\partial \dot{A}^i}\>,\quad
\pi_i = \frac{\partial \mathcal{L}}{\partial \dot{P}^{0i}}\>,\quad
\pi_{ij}=\frac{\partial \mathcal{L}}{\partial \dot{P}^{ij}}\>,
\label{momenta1a}
\end{equation}
we find from (\ref{Lagrangian-density2}) the following primary constraints
\begin{eqnarray}
&&\Delta^1 = \Pi_0^A \approx 0\>,\quad
\Delta^2_i = \Pi_i^A \approx 0\>,\quad \Delta^3_i = \pi_i-A_i \approx 0\>,\quad
\Delta^4_{ij} = \pi_{ij}\approx 0\>,
\label{primary-constraints}
\end{eqnarray}
defining a constraint surface on phase space (on which they vanish weakly in
Dirac's terminology \cite{constraint-analysis}).
Following Dirac's method,
we introduce Lagrange multiplier fields $\lambda^1$, $\lambda^2_i$, $\lambda^3_i$
and $\lambda_{ij}^4$ and define the extended Hamiltonian
$H_E = \int d^3x\,\mathcal{H}_E$, with
\begin{eqnarray}
\mathcal{H}_E&=& (\partial_i A_0) P^{i0}+(\partial_i A_j) P^{ij} + V(P,Q)+A_\mu J^\mu + \lambda^1\Delta^1+\lambda^2_i\Delta^{2i}+\lambda^3_i\Delta^{3i}+\lambda^4_{ij}\Delta^{4ij}\>.
\end{eqnarray}
Imposing that the time evolution of the constraints, $\dot{\Delta}^k=\{\Delta^k,H_E\}$,
vanish weakly yields the conditions 
\begin{eqnarray}
\Sigma^1 &=& \partial_i P^{i0}-J^0 \approx 0\>,
\label{Sigma1}\\
\Sigma^2_i &=& \lambda^3_i-\partial_m P^{mi}-J_i \approx 0\>,
\label{Sigma2}\\
\Sigma^3_i &=& \partial^i A_0+ V_P(P,Q) P^{0i}+V_Q(P,Q)\tilde{P}^{0i}-\lambda^{2}_i \approx 0\>,
\label{Sigma3}\\
\Sigma^4_{ij} &=& -(\partial_i A_j-\partial_j A_i+ V_P(P,Q)P_{ij}+ V_Q(P,Q)\tilde{P}_{ij}) \approx 0\>.
\label{Sigma4}
\end{eqnarray}
where $V_P$ and $V_Q$ indicate the partial derivatives
$\frac{\partial V}{\partial P}$ and $\frac{\partial V}{\partial Q}$,
respectively.
The conditions $\Sigma^2_i\approx0$ and $\Sigma^3_i\approx0$ can be used to
fix the Lagrange multipliers $\lambda^3_i$ and $\lambda^2_i$, respectively,
while Eqs.\ (\ref{Sigma1}) and (\ref{Sigma4}) define additional, secondary constraints.
It can be checked that imposing that the time evolution of $\Sigma^1$ and $\Sigma^4_{ij}$ vanish weakly does not produce any more constraints.
Dirac's method therefore terminates at this point, and we end up with the constraints: $\Delta^1,\Delta^2_i,\Delta^3_i,\Delta^4_{ij},\Sigma^1,\Sigma^4_{ij}$.

In order to split the constraints in first- and second-class constraints,
let's define the new set $\Theta^a$, $a=1,2,...,6$ as
\begin{eqnarray}
\Theta^1 &=& \Delta^1 = \Pi_0^A \approx 0\>, 
\label{Theta-1}\\
\Theta^2 &=& \Sigma^1 + \partial^i\Delta^2_i = \partial_i P^{i0} + \partial^i\Pi_i^A - J^0 \approx 0\>, \\
\Theta^3_i &=& \Delta^2_i = \Pi_i^A \approx 0\>,\\
\Theta^4_i &=& \Delta^3_i = \pi_i-A_i \approx 0\>,\\
\Theta^5_{ij} &=& \Delta^4_{ij} = \pi_{ij} \approx 0\>,\\
\Theta^6_{ij} &=& -\Sigma^4_{ij} = (\partial_i A_j-\partial_j A_i)+ V_P P_{ij}+ V_Q\tilde{P}_{ij} \approx 0\>.
\label{Theta-6}
\end{eqnarray} 
It can be easily proved that $\Theta^1$ and $\Theta^2$ commute with all the constraints,
i.e., they are first-class constraints.
The determinant of the matrix of Poisson brackets of the remaining constraints
$\Theta^a$ ($a = 3,4,5,6$) is
\begin{eqnarray}
\left|\left\{\Theta^a(x),\Theta^b(y)\right\}_{x^0=y^0}\right| &=&
V_P^2 S^2\delta^3(\vec{x}-\vec{y})\>.
\label{det-constraints}
\end{eqnarray}
where we defined
\begin{eqnarray}
S &=&\left(V_{PP}V_{QQ}-V_{PQ}^2\right)\left(H^2D^2 - Q^2\right)+ V_P\left(D^2 V_{QQ} + H^2 V_{PP} - 2 Q V_{PQ}\right) + V_P^2\>,
\label{S}
\end{eqnarray}
with
\begin{equation}
D = |\vec D|
\qquad\mbox{and}\qquad
H = |\vec H|\>.
\end{equation}
Since the right-hand side of Eq.\ (\ref{det-constraints}) is generally nonzero,
we conclude that $\{\Theta^3_a, \Theta^4_a, \Theta^5_{ab}, \Theta^6_{ab}\}$
form a second-class constraint set.
Thus,
as our model contains $6 + 4 = 10$ variables in the coordinate space,
with 2 first-class and 12 second-class constraints,
the number of phase space degrees of freedom is  
\begin{equation}
\#\mbox{d.o.f.} = 2 \times 10 - 2\times2 - 12 = 4\>,
\end{equation}
as expected.

In this work, we are interested in investigating global stability of
the model as well as the existence of local minima.
In order to do so, we first fix the gauge degrees of freedom that
are generated by the first-class constraints $\Theta^1$ and $\Theta^2$
by adding the gauge-fixing constraints
\begin{equation}
\chi^1 = A^0, \qquad \chi^2 = \partial_i A^i\>,
\label{gauge-fixing-constraints}
\end{equation} 
which convert $\Theta^1$ and $\Theta^2$ to second-class.
By setting all the constraints strongly equal to zero,
the effective Hamiltonian density can be expressed as
\begin{eqnarray}
\mathcal{H}_{\mbox{\scriptsize eff}} &=& (\partial_i A_j) P^{ij} + V(P,Q) \nonumber \\
&=&-\frac{1}{2}(V_P P_{ij}P^{ij} + V_Q P^{ij}\tilde{P}_{ij}) + V(P,Q) + A_i J^i\>,
\end{eqnarray}
where we have employed the constraint $\Theta^6_{ij}=0$ to obtain the second line. 
The remaining local degrees of freedom are contained in $P_{ij}$ and $P^{i0}$,
subject to the constraints (initial conditions)
\begin{equation}
\partial_i P^{i0} = J^0
\qquad\mbox{and}\qquad
\partial_k \epsilon^{ijk} (V_P P_{ij}+ V_Q\tilde{P}_{ij}) = 0 \>,
\label{initial-conditions1}
\end{equation}
which arise from the constraints $\Theta^2$ and $\Theta^6_{ij}$.
In terms of the fields $\vec D$ and $\vec H$,
the effective Hamiltonian density becomes
\begin{equation}
\mathcal{H}_{\mbox{\scriptsize eff}} = -H^2 V_P - QV_Q + V(P,Q) + A_i J^i,
\label{effective-Hamiltonian}
\end{equation}
while the initial conditions (\ref{initial-conditions1}) can be seen to correspond
to the Gauss' laws (\ref{Gauss-D}) and (\ref{Gauss-B})
\begin{equation}
\vec\nabla \cdot \vec D = J^0
\qquad\mbox{and}\qquad
\vec\nabla \cdot \vec B = 0\>,
\end{equation} 
where we identified the magnetic field
\begin{equation}
B^k = \epsilon^{ijk}\partial_i A_j
\end{equation}
and used constraint $\Theta^6_{ij}$.

As it turns out,
the effective Hamiltonian (\ref{effective-Hamiltonian})
corresponds exactly to the 00~component of the energy-momentum tensor
associated to the Lagrangian defined by Eq.\ (\ref{Lagrangian-density})
(see \cite{Plebanski}),
which serves as a nice check on the consistency of the formalism.

\section{Stability and local minima}
\label{sec:stability}
In this work we choose to limit our attention to potentials that lead to
an effective Hamiltonian (\ref{effective-Hamiltonian}) that is globally bounded
from below, in order to assure stability of the model.
In this analysis we will choose the external current $\vec J$ to be zero.
We will assume the potential can be expanded as power series
of monomials in $P$ and $Q$:
\begin{equation}
V(P,Q)=-\sum_{m,n\ge0} C_{mn} P^m Q^n \>,
\label{VPQ}
\end{equation}
where $C_{mn}$ are constant coefficients of mass dimension
$4(1-m-n)$ (note that $P$, $Q$ and $V(P,Q)$ have all mass dimension 4).
Moreover, in the (classical) analysis in this paper we will assume,
for simplicity, that the expansion (\ref{VPQ}) is convergent for all
values of $P$ and $Q$.
In order to have a good control of the stability we will
separate the dependence of the potential on $P$ and $Q$ and assume
the special form
\begin{equation}
V(P,Q) = V_1(P) + V_2(Q) \>.
\label{VPQ-special}
\end{equation}
so that $V_{PQ}$ is identically equal to zero.
We can expand
\begin{eqnarray}
V_1(P) &=& -\sum_{k\ge1}\alpha_k P^k\>,\nonumber\\
V_2(Q) &=& -\sum_{m\ge1}\beta_m Q^m
\label{VPQ-special2}
\end{eqnarray}
where $\alpha_m$ and $\beta_m$ are constant coefficients
of mass dimension $-4(m-1)$.
The corresponding effective Hamiltonian is then given by
\begin{equation}
\mathcal{H}_{\mbox{\scriptsize eff}} = \frac{1}{2}\sum_{k\ge1}\alpha_k P^{k-1}[(2k-1)H^2 + D^2]+\sum_{m\ge1}(m-1)\beta_{m} Q^{m}\>.
\label{Hamiltonian-eff}
\end{equation}
A sufficient (but not necessary) condition for
$\mathcal{H}_{\mbox{\scriptsize eff}}$ to be bounded from below
is to select all $\alpha_{k}$ and $\beta_{m}$ to be zero
for $k$ even and $m$ odd,
and non-negative for $k$ odd and $m$ even.
As it turns out, such a strong restriction does not lead to nontrivial
local minima of the effective Hamiltonian,
something we want to investigate in this work. 
Fortunately, there are many potentials of the form
 given by Eqs.\ \eqref{VPQ-special}
and \eqref{VPQ-special2}
where this restriction is relaxed,
but which nevertheless are associated with an effective Hamiltonian that is
bounded from below.

Let us now consider the conditions determining 
stationary points of the effective Hamiltonian density:
\begin{equation}
\frac{\partial\mathcal{H}_{\mbox{\scriptsize eff}}}{\partial D_i} = 0
\qquad\mbox{and}\qquad
\frac{\partial \mathcal{H}_{\mbox{\scriptsize eff}}}{\partial H_i} = 0\>.
\qquad i=1,2,3.
\label{stationary-H1}
\end{equation}
Noting that $\mathcal{H}_{\mbox{\scriptsize eff}}$ can be taken to depend
on the independent quantities $H^2$, $D^2$ and $Q$,
conditions (\ref{stationary-H1}) can be written as
\begin{equation}
2H_i \mathcal{H}_{H^2} - D_i \mathcal{H}_Q = 0
\,\,,\,\,
2D_i \mathcal{H}_{D^2} - H_i \mathcal{H}_Q = 0\>,
\quad i=1,2,3,
\label{stationary-H2}
\end{equation}
(suppressing, for simplicity, the subscript "eff" on $\mathcal{H}$),
where, in terms of the potential $V$
\begin{eqnarray}
\mathcal{H}_{H^2} &=& -\frac{1}{2}(V_P + H^2 V_{PP} + QV_{PQ}) = -\frac{1}{2}(V_P + H^2 V_{PP})\>,\\
\mathcal{H}_{D^2} &=& -\frac{1}{2}(V_P - H^2 V_{PP} - QV_{PQ}) = -\frac{1}{2}(V_P - H^2 V_{PP})\>,\\
\mathcal{H}_Q &=& -H^2 V_{PQ} - QV_{QQ} = -QV_{QQ}\>.
\label{conditions1abc}
\end{eqnarray}
In order to verify whether a stationary point corresponds to a local minimum of $\mathcal{H}$,
we need the Hessian $6\times6$-matrix
\begin{equation}
\frac{\partial^2 \mathcal{H}}{\partial X_i\partial X_j}\>,
\qquad X_i=D_i,H_i.
\label{Hessian1}
\end{equation}
This matrix can be written in terms of $3\times3$ blocks as

\begin{equation}
\left(
\begin{array}{cc}
2\mathcal{H}_{D^2}\delta_{ij} + 4\mathcal{H}_{D^2D^2}D_i D_j + \mathcal{H}_{QQ} H_i H_j
& \mathcal{H}_Q\delta_{ij} + (\mathcal{H}_{QQ}+\mathcal{H}_{D^2H^2} )D_i H_j \\
\mathcal{H}_Q\delta_{ij}+ (\mathcal{H}_{QQ}+\mathcal{H}_{D^2H^2} )H_i D_j
& 2\mathcal{H}_{H^2} \delta_{ij} + 4\mathcal{H}_{H^2H^2} H_iH_j + H_{QQ} D_i D_j
\end{array}
\right)\>,
\label{Hessian2}
\end{equation}

where $i,j=1,2,3$ and
\begin{equation}
\mathcal{H}_{X}=\frac{\partial\mathcal{H}}{\partial X}\>,\qquad
\mathcal{H}_{XY}=\frac{\partial^2\mathcal{H}}{\partial X\partial Y }\>,\qquad
(X,Y = H^2,D^2,Q)\>.
\end{equation}
Let us now consider various possible solutions of conditions (\ref{stationary-H2}).
\subsection*{Case 1: $\vec D = \vec H = 0$}
In this case the conditions \eqref{stationary-H2} for a stationary point
(corresponding to the ``canonical vacuum'') are evidently satisfied.
For $\vec{D} = \vec{H} = 0$,
the Hessian (\ref{Hessian2}) then reduces to 
\begin{equation}
\left(
\begin{array}{cc}
-V_P\delta_{ij} & 0 \\
0 & -V_P\delta_{ij}
\end{array}
\right)
\end{equation}
which is proportional to the $6\times 6$ identity matrix.
We therefore conclude that $\vec D = \vec H = 0$ is a local minimum of
$\mathcal{H}$ if the potential satisfies $V_P < 0$ in the vacuum.
For potentials of the form (\ref{VPQ-special}) this means that we need
$\alpha_1 > 0$.

\subsection*{Case 2: $\vec D = 0$, $\vec H \ne 0$}

Now we have $Q = 0$, $P = \frac{1}{2}H^2$,
and the conditions (\ref{stationary-H2}) become
\begin{equation}
V_P + 2PV_{PP} = 0\>.
\label{conditions-case2}
\end{equation}
In fact, condition (\ref{conditions-case2}) is equivalent to the condition
$S = 0$ (where $S$ is defined by Eq.(\ref{S})). 

A particular example is given by the following polynomial form for $V(P)$:
\begin{equation}
V(P,Q) = V_1(P) = -\alpha_1 P - \alpha_3 P^3 - \alpha_5 P^5\>.
\label{polynomial-V}
\end{equation}
As is shown in the Appendix,
the associated effective Hamiltonian is bounded from below provided we choose:
\begin{equation}
\alpha_1 > 0 \>,\qquad
\alpha_3 > -\sqrt{\frac{20}{9}\alpha_1\alpha_5}
\qquad\mbox{and}\qquad
\alpha_5 > 0 \>.
\label{conditions_bounded-from-below}
\end{equation}
It can be verified that the configurations
\begin{equation}
D^2 = 0
\qquad\mbox{and}\qquad
H^2 = 2\sqrt{\frac{\sqrt{5\alpha_3^2 - 4\alpha_1\alpha_5} - \sqrt{5}\alpha_3}{6\sqrt{5}\alpha_5}}
\label{criticalpoint-case2}
\end{equation}
satisfy the critical-point conditions (\ref{conditions-case2})
if we take
\begin{equation}
\alpha_3 < - \sqrt{\frac{4\alpha_1\alpha_5}{5}}
\label{alpha_3-conditions2}
\end{equation}
together with relation (\ref{conditions_bounded-from-below}).
At the critical point the Hessian matrix (\ref{Hessian2}) takes the form
\begin{equation}
\left(
\begin{array}{cc}
2\mathcal{H}_{D^2}\delta_{ij} 
& 0_{3\times3} \\
0_{3\times3} & 4\mathcal{H}_{H^2H^2} H_iH_j
\end{array} \right)\qquad i,j = 1,2,3
\end{equation}
which has a triple eigenvalue given by
\begin{equation}
2\mathcal{H}_{D^2} =
\frac{4}{9}\left(-\frac{\alpha_3^2}{\alpha_5}+\frac{\alpha_3}{\alpha_5}\sqrt{\alpha_3^2-\frac{4\alpha_1\alpha_5}{5}} + 4\alpha_1\right) \>,
\label{triple-eigenvalue-case-2}
\end{equation}
a simple eigenvalue
\begin{equation}
4\mathcal{H}_{H^2H^2} H^2 =
10\left(\frac{\alpha_3^2}{\alpha_5}-\frac{\alpha_3}{\alpha_5}\sqrt{\alpha_3^2-\frac{4\alpha_1\alpha_5}{5}} - \frac{4}{5}\alpha_1\right)\>,
\label{simple-eigenvalue-case-2}
\end{equation}
as well as a double eigenvalue zero.
From conditions (\ref{conditions_bounded-from-below}) and
(\ref{alpha_3-conditions2}) it follows that
the eigenvalues (\ref{triple-eigenvalue-case-2}) and
(\ref{simple-eigenvalue-case-2}) are both positive,
while the double zero eigenvalue has its origin in the fact
that there are two spontaneously broken Lorentz generators in this case
(corresponding to the rotations that rotate $\vec H$).
\begin{figure}
\centering
\includegraphics[scale=0.75]{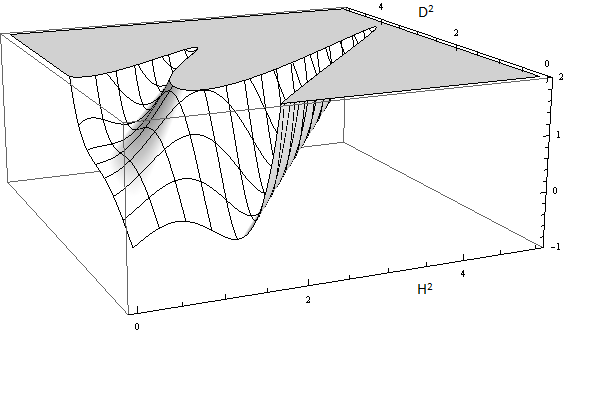}
\caption{The value  of $\mathcal{H}_{\mbox{\scriptsize eff}}$ as a function of $H^2$ and $D^2$ for
the parameter values $\alpha_1 = \alpha_5 = 1$, $\alpha_3 = -1.25$.
For clarity, the vertical range is cut off at the (arbitrary) value 2.}
\label{fig:example}
\end{figure}
As an example, we have plotted in Fig.\ \ref{fig:example} the
effective Hamiltonian as a function of $H^2$ and $D^2$ for the values
$\alpha_1 = \alpha_5 = 1$, $\alpha_3 = -1.25$.
We see that there is a (type 1) local minimum at $H^2 = D^2 = 0$,
but the absolute minimum of the Hamiltonian density is of type 2,
occurring for $D^2 = 0$, $H^2 \approx 1.2$.

Note that at the critical point $V_P = -\mathcal{H}_{D^2} \ne 0$,
$V_Q = 0$, so that, 
from relations (\ref{constitutive-relations3a}), we have at the critical point
\begin{equation}
\vec E = 0\>,\qquad
\vec B = \mathcal{H}_{D^2} \vec H\>,
\end{equation}
with $|\vec H|$ fixed by (\ref{criticalpoint-case2}). 
Thus, the vacuum has a background Lorentz-violating $\vec B$ field,
which can be probed directly by coupling it to a suitable current.

It is worth to mention that while at the critical point
$V_P \ne 0$, the factor in square
brackets on the right-hand side of Eq.\ (\ref{det-constraints}) vanishes.
Thus the determinant of the second-class constraints has a double zero at the
critical point.

One might be tempted to associate the eigenvalues of the Hessian
with (half) the square of the velocities of propagation of the physical modes,
as is the case for (linear) Maxwell theory.
Unfortunately, in order to extract the dynamics from the effective
Hamiltonian (\ref{effective-Hamiltonian}) one needs to use Dirac brackets
$\{\ ,\ \}_D$, which are not straightforward to obtain.
Instead, the dynamics around the vacuum defined by the solution
(\ref{criticalpoint-case2}) can be extracted from the Maxwell equations
(\ref{Gauss-D})--(\ref{Faraday}).
To this effect,
we linearize the model around the critical point by expanding
$\vec{D} = \vec{d}$, $\vec{H} = \vec{H}_0 + \vec{h}$, where
$\vec H_0$ satisfies condition (\ref{criticalpoint-case2}).
 We will assume in the following that $\vec{H}_0$ is a uniform field,
that is, it is independent of the position in space.
Using the constitutive relations (\ref{constitutive-relations3a}),
we can express $\vec E$ and $\vec B$ in terms of $\vec{d}$ and $\vec{h}$. 
Expanding the Maxwell equations (\ref{Gauss-D})--(\ref{Faraday})
to linear order in the fluctuations ($\vec{d},\vec{h}$),
one obtains
\begin{eqnarray}
\vec{\nabla}\cdot \vec{d} &=& J^0\>,
\label{linearized-Maxwell-case2_a} \\
\vec{\nabla}\times \vec{h}-\frac{\partial\vec{d}}{\partial t} &=& \vec{J}\>,
\label{linearized-Maxwell-case2_b} \\
\vec{\nabla}\cdot\vec{h}_\bot &=& 0\>,
\label{linearized-Maxwell-case2_c} \\
\vec{\nabla}\times \vec{d} + \frac{\partial{\vec{h}_\bot}}{\partial t} &=& 0
\label{linearized-Maxwell-case2_d}
\end{eqnarray}
where
\begin{equation}
\vec{h}_\bot = \vec{h}-\frac{(\vec{h}\cdot\vec{H}_0)}{\vec{H}^2_0}\vec{H}_0\>.
\label{hperpen}
\end{equation}

Let us take $J^0  = \vec J = 0$ and look for plane-wave solutions
of the form 
\begin{eqnarray}
\vec{h}_\bot &=& \vec{h}_{\bot 0} e^{i(k\cdot r-\omega t)}\>, \\
\vec{h}_\| &=& \phi_0 \vec{H}_0 e^{i(k\cdot r-\omega t)}\>, \\
\vec{d} &=& \vec{d}_0 e^{i(k\cdot r-\omega t)},
\end{eqnarray}
where $\vec{h} =  \vec{h}_\bot + \vec{h}_\|$,
$\phi_0$ is a constant,
$\vec{d}_0$ and $\vec{h}_{\bot 0}$ are constant vectors with
$\vec{h}_{\bot 0} \cdot \vec H_0 = 0$. It is easy to check that
Eqs.\ (\ref{linearized-Maxwell-case2_a})--(\ref{linearized-Maxwell-case2_d})
imply that $\phi_0 = 0$,
$\vec{h}_{\bot 0}$ is perpendicular to the plane $M(\vec k,\vec H_0)$
spanned by $\vec k$ and $\vec H_0$,
while $\vec d_0$ is in $M(\vec k,\vec H_0)$, with $\vec d_0 \perp \vec k$.
Thus, at linearized level there is only one propagating mode rather than
the usual two.
This should not come as a surprise, considering the fact that at the
critical point, the determinant of the second-class constraints has a double zero,
indicating the loss of two of the four phase-space degrees of freedom.
The dispersion relation of the remaining mode is the usual one
\begin{equation}
\omega^2 = \vec k^2\>.
\end{equation}   
showing that it behaves as a usual massless field.
This is consistent with the fact that the perpendicular fluctuations
$\vec{h}_{\bot}$ correspond to a Nambu-Goldstone mode associated with a broken
Lorentz generator (corresponding to rotations of $\vec{H}_0$, see above).

For the dynamics of the remaining mode $\vec h_\|$,
it is unfortunately not possible to use the linearized equations of motion,
because they yield an empty equation at the critical point.
For this reason, it is necessary to resort to the full nonlinear 
equations of motion.
Such an analysis has been done for general Pleba\'nski models in
\cite{paper2}.
The most important conclusion of that work is that the dynamics of the
$\vec h_\|$ mode can develop a degeneracy on/near hypersurfaces defined
by $S = 0$ (with $S$ defined by Eq.\ (\ref{S})),
involving shock-wave-like and/or superluminal motion.
As noted below Eq.\ \eqref{conditions-case2},
the condition $S = 0$ is in fact satisfied at the case 2 vacuum,
and therefore the dynamics of the $\vec h_\|$ mode can be expected to
involve the type of behavior found in \cite{paper2}.

In a recent work \cite{SchellstedePerlickLammerzahl},
Schellstede et.\ al.\ analyzed nonlinear vacuum electrodynamics
in the eikonal approximation for arbitrary potentials and
managed to derive conditions for causality to hold.
They found that causality holds if $V_P \ne 0$ and the conditions
\begin{eqnarray}
\frac{V_{PP}}{V_P} &&\ge 0\>, \label{causality-conditions_a}
\\
\frac{V_{QQ}}{V_P} &&\ge 0\>, \\
V_{PP} V_{QQ} - V_{PQ}^2 &&\ge 0\>, \\
\frac{V_{PP}\left(\sqrt{P^2 + Q^2} - P\right) + P V_{QQ} - 2Q V_{PQ}}{V_P} &&< 1
\label{causality-conditions_d}
\end{eqnarray}
are satisfied for all possible values of $P$ and $Q$
(here we translated the expressions in \cite{SchellstedePerlickLammerzahl}
to our somewhat different conventions).
For the potential (\ref{polynomial-V}) this is clearly not the case.
For instance, the vacuum configuration $Q = 0$, $P = H^2/2 > 0$,
satisfies condition (\ref{conditions-case2}),
which implies that condition (\ref{causality-conditions_a})
 should be violated in the vacuum itself.
This conclusion is clearly consistent with the conclusions
reached in \cite{paper2} as discussed above.

\subsection*{Case 3: $\vec D \ne 0$, $\vec H = 0$}

To analyze this case we use again the potential (\ref{polynomial-V}) subject
to the conditions (\ref{conditions_bounded-from-below}).
We have, by assumption, $H^2=Q=0$,
and thus the conditions (\ref{stationary-H2}) reduce to $V_P = 0$.
Imposing positivity of the eigenvalues of the Hessian yields
\begin{equation}
\alpha_3 \le -\sqrt{\frac{20}{9}\alpha_1\alpha_5}
\end{equation}
at the critical point,
which is in contradiction with conditions (\ref{conditions_bounded-from-below}).
It follows that there are no stationary points of this type if we impose
that the effective Hamiltonian be bounded from below.
It is possible that there are other forms of the potential that lead to
an effective Hamiltonian that is bounded from below and has local minima,
but we will not pursue them in this work.

\subsection*{Case 4: $\vec D \ne 0$, $\vec H \ne 0$}

The conditions (\ref{stationary-H2}) now imply for the stationary points
\begin{equation}
V_P = 0\>,\quad
H^2 V_{PP} + QV_{PQ} = 0
\quad\mbox{and}\quad
H^2 V_{PQ} + QV_{QQ} = 0\>.
\label{stationary-conditions-4}
\end{equation}
One readily verifies that the conditions (\ref{stationary-conditions-4})
together with the condition $\vec H \ne 0$ imply that the quantity $S$
defined in (\ref{S}) vanishes.
To illustrate this case we will take a different example than
for the previous cases, given by the potential
\begin{equation}
V(P,Q)=-\alpha'_3 (P-p)^3 - \beta'_4\left(\frac{Q^4}{3} - 2q^2Q^2 - q^4\right)
\label{potential-case-4}
\end{equation}
where $\alpha'_3$, $\beta'_4$, $p$ and $q$ are constants.
It is of the general form given by Eqs.\ (\ref{VPQ-special}) and
(\ref{VPQ-special2}).
The effective Hamiltonian (\ref{Hamiltonian-eff}) becomes
\begin{equation}
\mathcal{H}_{\mbox{\scriptsize eff}} =
\frac{\alpha'_3}{2} (P-p)^2 (5H^2 + D^2 + 2p) + \beta'_4(Q^2 - q^2)^2\>,
\label{H-eff-4}
\end{equation}
which is evidently bounded from below if we take $\alpha'_3$, $\beta'_4$
and $p$ positive.
The conditions (\ref{stationary-conditions-4}) yield
\begin{equation}
P = p
\qquad\mbox{and}\qquad
Q = \pm q\>.
\label{solution-4}
\end{equation}
For these configurations $\mathcal{H}_{\mbox{\scriptsize eff}}$ in
\eqref{H-eff-4} takes its absolute minimum value 0.
Eqs.\ (\ref{solution-4}) are solved by taking
\begin{equation}
H^2 = 2p + D^2
\qquad\mbox{and}\qquad
\cos\theta = \frac{\mp q}{\sqrt{D^2(2p + D^2)}}\>,
\label{solution-4HD}
\end{equation}
expressing $H^2$ and the angle $\theta$ between $\vec H$ and $\vec D$
in terms of $D^2$.
From the second equation of (\ref{solution-4HD})
we easily derive that $D^2$
can take any value not smaller than $\sqrt{p^2 + q^2} - p$.
For the special case $q = 0$, $\vec{H}$ and $\vec{D}$ are always perpendicular.

This case is rather different from the previous cases in that the conditions
(\ref{solution-4}) that minimize the effective Hamiltonian are Lorentz invariant.
A straightforward calculation shows that the Hessian matrix (\ref{Hessian2})
has two positive eigenvalues
$6\alpha'_3 H^2(H^2+D^2)$ and $8\beta'_4 Q^2(H^2+D^2)$ at the minimum,
as well as a quadruple zero eigenvalue.
(These correspond to Nambu-Goldstone (NG) modes, associated to
four spontaneously broken Lorentz generators.
The remaining two combinations of boosts and rotations leave
any configuration satisfying (\ref{solution-4HD}) invariant,
and therefore they do not generate NG modes.)
It was already shown by Escobar and Urrutia \cite{EscobarUrrutia}
who studied spontaneous Lorentz violation in nonlinear electromagnetism
that any combination of nonzero vectors $\vec E$ and $\vec B$
gives rise to four NG modes. 

As an aside,
one may wonder about the fact that the classification defining the cases
2, 3 and 4 is not Lorentz covariant.
The reason is that,
even though the potential $V$ is a Lorentz-invariant quantity,
the effective Hamiltonian $\mathcal{H}_{\mbox{\scriptsize eff}}$ is not.
As a consequence, the distinction of vacua in terms of four classes,
while not Lorentz covariant, is nonetheless relevant,
as the vacua in different cases are not equivalent.
The spontaneous symmetry breaking is not realized in the same way in the
different cases because the vacua preserve different symmetries.
A good way to see this is by looking at the number of NG modes,
which is different for the cases 2 and 4.

It is worth to mention that in a minimum
defined by conditions (\ref{solution-4HD})
$V_P = 0$ and $V_Q = 8\beta'_4 q^3/3$,
so that from the constitutive relations (\ref{constitutive-relations3a}) we find
\begin{equation}
\vec{E} = -\frac{8}{3}\beta'_4 q^3 \vec{H}
\qquad\mbox{and}\qquad
\vec{B} = \frac{8}{3}\beta'_4 q^3 \vec{D}
\label{E,B-case4}
\end{equation}
for the values of the electric field and the magnetic induction.

In order to obtain the equations of motion,
we proceed like we did for case~2 and write
\begin{equation}
\vec D = \vec D_0 + \vec d\>,\quad
\vec H = \vec H_0 + \vec h\>,\quad
\vec E = \vec E_0 + \vec e\>,\quad
\vec B = \vec B_0 + \vec b\>,
\label{fluctuations}
\end{equation}
where the lower-case letters indicate the fluctuations around the minimum.
By substituting (\ref{fluctuations}) in the Maxwell equations,
and expressing $\vec e$ and $\vec b$ in terms of $\vec d$ and $\vec h$
through the constitutive relations (\ref{constitutive-relations3a}),
one can obtain the time development of each of the components of
$\vec d$ and $\vec h$ as a function of the space derivatives.
While in the case 2 vacuum one of the propagating modes could be described
by the linearized equations of motion
(\ref{linearized-Maxwell-case2_a})--(\ref{linearized-Maxwell-case2_d}),
in the case 4 vacuum it is necessary for both modes to analyze
the full nonlinear equations of motion.
As mentioned below Eq.\ (\ref{stationary-conditions-4}),
the conditions defining the case 4 vacuum imply that the quantity $S$
is zero.
Therefore, following the analysis in \cite{paper2},
just like as we saw in the case of the $h_\|$ mode in the case 2 vacuum,
the fluctuations (\ref{fluctuations}) can be expected to involve
shock wave motion and superluminal propagation.

It is interesting to note from Eq.\ (\ref{E,B-case4})
that $\vec E$ and $\vec B$ vanish when we take $q = 0$.
Correspondingly, in that case we have $V_P = V_Q = 0$ in the
minimum defined by conditions (\ref{solution-4HD}),
and thus the potential $V(P,Q)$ itself is stationary.
Note, however, that $V(P,Q)$ does not take an extremal value,
even though $\mathcal{H}_{\mbox{\scriptsize eff}}$ does.

 Finally, let us take a look at the issue of causality for case~4.
One readily sees that, unfortunately, 
conditions (\ref{causality-conditions_a})--(\ref{causality-conditions_d})
cannot be used because $V_P = 0$ in the vacuum.

\section{Phenomenology}
\label{sec:phenomenology}

In the previous section we have seen that there exist potentials
corresponding to effective Hamiltonians that are bounded from below
with local minima leading to spontaneous breaking of Lorentz invariance,
through nonzero expectation values of $P_{\mu\nu}$ and,
possibly, $F_{\mu\nu}$.
Let us stress again the point made in the introduction,
that the nonlinear electrodynamics we envision in this work
evidently cannot
 be taken to correspond to a small correction of
Maxwell electrodynamics, as it involves large modifications
in the nontrivial vacua identified in the previous section.
Instead, we envision that it corresponds
to some other (so far unobserved) $U(1)$ gauge field.
One straightforward way in which any nonzero expectation value
of $F_{\mu\nu}$ could be detected is through the minimal
coupling to an external (charged) current as included 
in the definition (\ref{Lagrangian-density}).
In that case any expectation value of $F_{\mu\nu}$ would
manifest itself through a charge-dependent Lorentz force acting
on particles.

Alternatively, nonminimal couplings may exist,
or might arise through quantum effects.
Consider, for instance, nonlinear electrodynamics coupled to a
Dirac spinor $\psi$.
Even if it is minimally coupled through the usual current
$J^\mu = \bar\psi\gamma^\mu\psi$,
a magnetic moment coupling
\begin{equation}
F_{\mu\nu}\bar\psi \sigma^{\mu\nu}\psi
\label{magnetic-moment-coupling}
\end{equation}
can be expected to arise in the one-loop effective action.
In vacua where $F_{\mu\nu}$ acquires a vacuum expectation value
$F_{0\,\mu\nu}$,
the term (\ref{magnetic-moment-coupling}) gives rise to
an Lorentz-violating coefficient $H^{\mu\nu} = -2F_0^{\mu\nu}$
in the QED sector of the SME \cite{SME}.
Similarly, the terms
$i F_{\mu\nu}\bar\psi\gamma^\mu\overset{\,\text{\scriptsize$\leftrightarrow$}}{D}{}^\nu\psi$
and
$i F_{\mu\nu}\bar\psi\gamma_5\gamma^\mu\overset{\,\text{\scriptsize$\leftrightarrow$}}{D}{}^\nu\psi$
yield contributions to the anti-symmetric parts of the SME
coefficients $c^{\mu\nu}$ and $d^{\mu\nu}$.

In this work we have restricted our attention to Minkowski space.
Novel results can be obtained by extending the scope to curved space.
For example,
we can couple the curvature in a covariant way to $F^{\mu\nu}$
through the terms
\begin{eqnarray}
&&R\left(a_1\, F^{\alpha\beta} F_{\alpha\beta}  +
a_2\, \epsilon_{\mu\nu\alpha\beta} F^{\mu\nu}F^{\alpha\beta}\right) +
b\, R_{\mu\nu} F^\mu{}_\alpha F^{\alpha\nu}  +
c\, R_{\mu\nu\alpha\beta} F^{\mu\nu}F^{\alpha\beta} \>,
\label{gravitational-coupling}
\end{eqnarray}
where $a_1$, $a_2$, $b$ and $c$ are constants.
In vacua where $F^{\mu\nu}$ has acquired a vacuum expectation value
$F_0^{\mu\nu}$, this amounts to an SME-type gravitational coupling of the form
$u R + s^{\mu\nu} R_{\mu\nu} + t^{\mu\nu\alpha\beta} R_{\mu\nu\alpha\beta}$,
where
\begin{eqnarray}
u &=& \left(a_1 + \tfrac{1}{4}b + \tfrac{5}{6}c\right) F_0^{\alpha\beta} F_{0\,\alpha\beta}+ a_2\,\epsilon_{\mu\nu\alpha\beta} F_0^{\mu\nu}F^{0\,\alpha\beta} \\
s^{\mu\nu} &=& (b + 2c)\left(F_0^\mu{}_\alpha F_0^{\alpha\nu} -
\tfrac{1}{4}F_0^{\alpha\beta} F_{0\,\alpha\beta} \eta^{\mu\nu}\right)\\
t^{\mu\nu\alpha\beta} &=& c \,\Bigl(F_0^{\mu\nu} F_0^{\alpha\beta} +
\tfrac{1}{24}\epsilon^{\mu\nu\alpha\beta}
\epsilon_{\lambda\rho\sigma\tau} F_0^{\lambda\rho}F_0^{\sigma\tau}+ \tfrac{1}{2}\left(F_0^{\rho\mu}F_{0\,\rho}{}^\beta \eta^{\nu\alpha}
 - F_0^{\rho\mu}F_{0\,\rho}{}^\alpha \eta^{\nu\beta}\right) \nonumber \\
 &&\qquad{} + \tfrac{1}{2}\left(
 + F_0^{\rho\nu}F_{0\,\rho}{}^\alpha \eta^{\mu\beta}
 - F_0^{\rho\nu}F_{0\,\rho}{}^\beta \eta^{\mu\alpha}\right) + \tfrac{1}{6} F_0^{\rho\sigma} F_{0\,\rho\sigma}
\left(\eta^{\mu\alpha}\eta^{\nu\beta} - \eta^{\mu\beta}\eta^{\nu\alpha}\right)\Bigr)
\end{eqnarray}
are constant tensors.
Here we have taken $s^{\mu\nu}$ to be traceless,
and $t^{\mu\nu\alpha\beta}$ with the symmetries of the Riemann tensor,
as defined in \cite{SMEgravity}.
However, note that fluctuations of $F^{\mu\nu}$ around $F_0^{\mu\nu}$
yield additional, time-de\-pen\-dent, Lorentz-violating contributions.
Very interesting is the case in vacua where $F_0^{\mu\nu}$ is zero,
which happens in the example presented for case 4
if we take $\beta'_4 = 0$ in Eq.\ (\ref{potential-case-4})
(the number of zero modes of the Hessian matrix
then becomes five rather than four).
Although there can then not be any SME-type contribution with constant 
coefficient depending on $F_0^{\mu\nu}$ (as this now vanishes),
there are couplings to the fluctuations that have a Lorentz-violating dynamics
themselves.
In other words,
Lorentz-violating effects are transmitted to the Riemann tensor
(or to the external current) in an indirect way,
presumably having the effect of suppression.
As an alternative to the couplings defined in Eq.\ (\ref{gravitational-coupling}),
one could substitute $F^{\mu\nu}$ by $P^{\mu\nu}$,
or even use mixed couplings involving one factor $F^{\mu\nu}$ and another
factor $P^{\mu\nu}$.

Let finish this section with a caveat.
As we have seen in the treatment of the cases 2 and 4 in the previous section,
it is still an open question whether there exist Lorentz-violating
vacua in which the dynamics of the fluctuations
of the $U(1)$ vector field is causal.
A vacuum with noncausal dynamics of the vector field would, of course,
only be acceptable if it does not give rise to detectable causality violations
for observable (Standard Model) fields.
Such a situation might arise if the coupling to Standard Model fields
is very weak.

\section{Conclusions and outlook}
\label{sec:conclusion}
In this work, we considered a class of potentials in nonlinear electrodynamics
in which the field strength acquires a non-zero vacuum expectation value (VEV),
using a first-order approach introduced by Pleba\'nski \cite{Plebanski}.
The spontaneous Lorentz-symmetry breaking that is triggered
this way constitutes an alternative to other models that
have been studied, such as Nambu's model or the bumblebee,
in which it is the vector potential that acquires the VEV.
The considerable advantage is that gauge invariance is maintained from the outset,
and that consistency requirements like stability can be guaranteed.

We performed a classical Hamiltonian analysis and,
employing Dirac's method, derived the constraints of the model.
They include both first- and second-class constraints,
the counting of which confirms that the model contains two degrees of freedom.
We then investigated the possible existence of local minima of the
effective Hamiltonian
(rather than of the potential, which makes an essential difference in this case).
We explicitly showed that there exist potentials
for which the corresponding Hamiltonian is globally bounded from below.
Local minima can be classified into four different types,
of which we presented examples.
In three of the cases Lorentz symmetry is spontaneously broken.
It turns out that, depending on the type of minimum,
the equations of motion can be singular, or partially singular,
at linear order in the fluctuations of the field.
If this happens, the matrix constructed of the Poisson brackets of
the second-class constraints becomes singular,
turning one or two of them into first-class constraints,
reducing the apparent number of degrees of freedom.
In that case, the dynamics at lowest order is obtained by including
terms at quadratic order in fluctuations. 

To the best of the authors' knowledge,
this is the first basic field-theoretical model
that exhibits spontaneous symmetry breaking of Lorentz symmetry,
while preserving gauge invariance,
where the Hamiltonian can be taken to be bounded from below.

As we have seen in section \ref{sec:stability}, a problematic
aspect of the physics of the Lorentz-breaking
vacua of type 2 and 4 is that one or more of the degrees of freedom
of the small fluctuations are subject to dynamics which
cannot be described by the linearized equations of motion.
Instead, it is necessary to resort to the full nonlinear equations
of motion, an issue that was studied in depth in Ref.\ \cite{paper2}.
As it turns out, there are regions in field space where the dynamics
of certain modes can develop degeneracies,
resulting in shock-wave-like and/or superluminal motion.
In particular, such behavior can be expected to be generated by the
equations of motion on and near the vacua of type 2 and 4,
both of which lie on such degenerate surfaces, defined by $S = 0$
(with $S$ defined by Eq.\ (\ref{S})).
This property of the Lorentz-breaking vacua evidently affects their
applicability as a realistic physical model.
Nevertheless, we could envision a situation in which the $U(1)$ field
either is very weakly coupled to the usual Standard Model degrees of
freedom, so that the degenerate behavior is undetectable,
or the field components are essentially frozen in a vacuum,
so that the only detectable effect is its vacuum value,
through a (weak) coupling to Standard Model fields and/or gravity,
as explained in section \ref{sec:phenomenology}.

We finish with a brief outlook on some open issues. 
We know that for potentials of type four there are four Nambu-Goldstone modes among the six phase-space degrees of freedom.
On the other hand, there must be four propagating degrees of freedom.
It is unclear if these are all NG modes,
or if one or two of them are auxiliary.
For a more elaborate discussion of this issue, see \cite{EscobarUrrutia}. 
In the literature there are a number of studies of light propagation in
nonlinear electrodynamics
(see, e.g., \cite{nonlinear-propagation1,nonlinear-propagation2}),
but not, as far as we know, around a configuration corresponding
to a nontrivial minimum of the effective Hamiltonian.

We restricted our study to potentials defined by
Eqs.\ (\ref{VPQ-special}).
However, it would be interesting to study the more general case as well.
For instance,
the potential $V(P,Q) = -\alpha (P + \beta Q^2)^3$,
with $\alpha$ and $\beta$ positive,
can be shown to produce a positive definite Hamiltonian.
Many other possibilities exist,
including nonpolynomial ones,
which may exhibit interesting properties not considered in this work.

As a final comment we note that, since gauge invariance is unbroken,
all degrees of freedom in nonlinear electrodynamics are necessarily massless.
It would be interesting to investigate
if some version of the Higgs mechanism can be applied that
would turn them massive.


\begin{acknowledgements}
We thank Leonor Cruzeiro,
Alan Kosteleck\'y and Luis Urrutia for useful comments and  discussions.
This work is supported in part by the Funda\c c\~ao para a Ci\^encia e
a Tecnologia of Portugal through grants SFRH\discretionary{/}{BSAB}{/BSAB}\discretionary{/}{150324}{/150324}/2019
and UID/FIS/00099/2019.
C.\ A.\ E.\ was supported by a UNAM-DGAPA postdoctoral fellowship
and the project PAPIIT No.\ IN111518.
\end{acknowledgements}

\appendix

\section{Conditions (\ref{conditions_bounded-from-below}) imply
effective Hamiltonian is bounded from below}
\label{app:bounded-from-below}
We rewrite the $P$-dependent part of the effective Hamiltonian 
\begin{eqnarray}
\mathcal{H}_{\mbox{\scriptsize eff}} =
&&2\alpha_1(H^2 + D^2) + 8(H^2 - D^2)^2\left(\alpha_3(5H^2 + D^2) + 4\alpha_5(9H^2 + D^2)(H^2 - D^2)^2\right)
\end{eqnarray}
by substituting $H^2 = D^2 + \delta$:
\begin{eqnarray}
\mathcal{H}_{\mbox{\scriptsize eff}} &=& D^2\left(4\alpha_1 + 48\alpha_3\delta^2 + 320\alpha_5\delta^4\right) + \delta\left(2\alpha_1 + 40\alpha_3\delta^2 + 288\alpha_5^4\right)\>.
\end{eqnarray}
With the condition
\begin{equation}
\alpha_3 > -\sqrt{\frac{20}{9} \alpha_1\alpha_5}
\label{cd2}
\end{equation} 
the first bracket is always positive for any value of $\delta$.
The argument is as follows:
the bracket defines a polynomial of second order in $\delta^2$ and,
with $\alpha_5$ positive,
this polynomial will be always positive either if it does not have real roots, 
which is true provided $9\alpha_3^2 < 20\alpha_1\alpha_5$,
or if $\alpha_3$ is positive, hence (\ref{cd2}).
The only possibility for the Hamiltonian to tend to $-\infty$ is
in either of the limits
$D^2 \to \infty$ or $\delta \to \pm\infty$.
If we fix $\delta$ and take the limit $D^2 \to \infty$
the condition (\ref{cd2}) guarantees that
$\mathcal{H}_{\mbox{\scriptsize eff}} \to \infty$.
If $\delta \to \infty$ (with any $D^2$) the dominant terms go to $+\infty$.
We note that for any $\delta < 0$
we must have $D^2 \geq |\delta|$, because $H^2 \geq 0$.
Moreover, the values of the polynomial in the first bracket
are always larger than the values of the polynomial in the second bracket
(for any $\delta$,
we can do the subtraction and calculate the discriminant to see that the result,
a new polynomial, does not have real roots, provided condition (\ref{cd2}) holds).
Thus, for $\delta \to -\infty$ the first term goes to plus infinity faster
than the second one goes to minus infinity
and therefore $\mathcal{H}_{\mbox{\scriptsize eff}}$ is bounded from below.

\end{document}